\documentstyle[aps]{revtex}

\def\bea{\begin{eqnarray}}
\def\eea{\end{eqnarray}}
\def\be{\begin{equation}}
\def\ee{\end{equation}}

\def\b{\beta}

\def\d{\delta}

\def\e{\epsilon}

\def\m\mu
\def\n{\nu}

\def\ph{\phi}

\def\x{\xi}

\def\p{\partial}

\begin{document}
 
\title{ Compactified D=11 Supermembranes\\ and Symplectic 
Non-Commutative\\ Gauge Theories}

\author{I. Mart\'{\i}n $^1$ \footnote{isbeliam@usb.ve 
;isbeliam@ic.ac.uk}  , J. Ovalle \footnote{jovalle@usb.ve} and A. Restuccia 
$^2$ \footnote{arestu@usb.ve}}
  
\address  {{\small Departamento de F\'{\i}sica, Universidad Sim\'on
Bol\'{\i}var, Venezuela} \\ {\small and} \\ {\small $^1$ Theoretical Physics 
Group, Imperial College , London University} \\ {\small $^2$ Department of Mathematics, 
King's College, London University.}}

\date{\today }

\maketitle

\begin{abstract}
    
It is shown that  a double compactified $D=11$ supermembrane  with non trivial wrapping may be 
formulated as a symplectic non-commutative gauge theory on the world volume.  
The symplectic non commutative structure is intrinsically obtained from the symplectic 2-form on the 
world volume defined by the minimal configuration of its hamiltonian. The 
gauge transformations on the symplectic fibration are
generated by the area preserving diffeomorphisms on the world volume. 
Geometrically, this gauge theory corresponds to a symplectic fibration over a 
compact Riemman surface with a symplectic connection.
\end{abstract}

\section{Introduction}

Noncommutative geometry in string theory with a
nonzero B-field \cite{con} has recently been discussed by several
authors \cite{dh}-\cite{vs}. The relation of noncommutative
Yang-Mills to Born-Infeld lagrangian was considered in \cite{hof},
\cite{pio} and general aspects of non-commutative gauge theories
have been discussed in \cite{howu}-\cite{giro}. In \cite{SW},the change of
variables from ordinary to noncommutative Yang-Mills was
explicitly found and the equivalence between the Born-Infeld
action for ordinary Yang-Mills in the presence of a B-field and of
a noncommutative Yang-Mills was proven. 

In this work, we follow a different approach. We relate the double compactified D=11 closed supermembrane \cite{BST} dual
\cite{PKT95}-\cite{mor} to a symplectic noncommutative gauge theory on
the world volume minimally coupled to seven scalar fields
representing the transverse coordinates to the brane.
 
We first show that there is a natural symplectic structure for the double compactified supermembrane with non trivial
wrapping on the target space. It is defined by the minimal configurations of the
hamiltonian \cite {mrt} ,\cite {mor}. In fact, the solutions when interpreted in terms of connection
1-forms over principle bundles satisfy the global condition 
\begin{equation}
^*F(A)=n.
\end{equation}

On the other hand, it is known that for a given a symplectic structure over a manifold there always exists a globally
defined deformation of the Poisson bracket. Moreover, even for 
Poisson manifolds it is possible to globally define a  Moyal bracket 
\cite {vey-kont} which leads to a non commutative geometry. Having this idea in mind, we show that the hamiltonian for the dual of the 
the double compactified supermembrane corresponds exactly to a Super-Maxwell theory of a symplectic 
connection on a symplectic fibration. The fibre being the space 
generated by transverse coordinates and its conjugate momenta to the brane 
in the target phase space. It is noticed that a deformation of the given 
geometrical structure in this theory will lead in a straightforward 
way to a non commutative ({\`{a}} la Moyal) gauge theory. The reformulation of the
compactified $D=11$ supermembrane dual in terms of noncommutative
gauge theories provides a different point of view to analyze
fundamental properties of the supermembrane as discussed in
\cite{dwln}, \cite{dwpp}. 

The steps taken in our formulation are as follow: we first construct 
the Hamiltonian for the doubly compactified supermembrane dual. The 
Hamiltonian minima are smooth configurations corresponding to $U(1)$ 
connections globally defined over the brane world volume. The 
curvature of these connections is a non degenerate 2-form that give  rise 
to a well defined symplectic structure.   
The second step in our construction is then to introduce symplectic 
connections with its covariant derivatives in the compactified 
directions. The Hamiltonian reduces then to an exact symplectic non 
commutative Super Maxwell theory interacting with scalar fields.

\section{Hamiltonian Formulation of the Double Compactified $D=11$ 
Supermembrane Dual}
We consider the compactified $D=11$ closed supermembrane dual obtained in \cite{PKT95}
and \cite{mor}. The bosonic part of its action is given by
\begin{equation}
S(\gamma ,X,A)=-\frac{1}{2}\int_{\Sigma x R} d^{3}\xi
\sqrt{-\gamma }\left( \gamma ^{ij}\partial _{i}X^{m}\partial
_{j}X^{n}\eta _{mn}+\frac{1}{2}\gamma ^{ij}\gamma
^{kl}F_{ik}^{r}F_{jl}^{r}-1\right) \label{db}
\end{equation}
where $X^m$, $m=1,...,D-q$ denote the maps from the world volume
$\Sigma$x$R$ to the target space
$M_{D-q}x\underbrace{S^1x...xS^1}_{q}$, $\Sigma$ been a compact
(closed) Riemann surface. $A^r_i$, $r=1,...,q$ denotes the
components of the $q\;U(1)$ connection 1-forms over $\Sigma$x$R$.
$\gamma^{ij}$ is the auxiliary metric. We will be interested in
the cases $q=1$ and $q=2$ the single and the double compactified
case. The case $q=0$ correspond to the supermembrane action over
$M_{11}$. The action (\ref{db}) for the $q=1$ case is dual to the
supermembrane with target space $M_{10}$x$S^1$, while the action
for $q=2$ is dual to the supermembrane with target space
$M_9$x$S^1$x$S^1$. The equivalence between the actions under
duality transformations is valid off-shell. The functional
integral formulations may be proven to be formally equivalent in
both cases.

To obtain the hamiltonian formulation of the theory, we consider
in the usual way, the ADM decomposition of the metric
\begin{equation}
\begin{tabular}{l}
$\gamma _{ab}=\beta _{ab} \hspace{1.2cm}\gamma
^{ab}=\beta^{ab}-N^{a}N^{b}N^{-2}$ \\ $\gamma
^{0a}=N^{a}N^{-2}\hspace{1.2cm}\gamma_{0a}=\beta_{ab}N^b$ \\
$\gamma_{00}=-N^2+\beta_{ab}N^aN^b\hspace{1.2cm}\gamma ^{00}=-N^{-2}$%
\end{tabular}
\label{adm}
\end{equation}
and define $\beta^{ab}$ by $\beta ^{ab}\beta _{bc}=\delta
_{c}^{a}$

The light-cone gauge fixing conditions are
\begin{equation}
X^{+}=P_{0}^{+}{\cal T} \hspace{1cm} P^{+}=P^{+}_0\sqrt{W}
\end{equation}
where $ {\cal T} $ is the time coordinate on the world volume and $W$ the determinant
of the metric over $ \Sigma $ introduced through the gauge fixing condition only.

After elimination of $X^-$ and $P^-$ one obtains \cite{mor} the hamiltonian density
\begin{equation}
{\cal H}=\frac{1}{2}\frac{1}{\sqrt{W}}\left( P^{M}P_{M}+\beta +\frac{1}{2}%
\beta\beta^{ac}\beta^{bd} F^{r}_{ab}F^{r}_{cd}\right)
-A_{0}^{r}\phi_r+\Lambda\phi \label{hamil}
\end{equation}
where $A^r_0$ and $\Lambda$ are the Lagrange multipliers
associated to the first class constraints
\begin{equation}
\phi_r\equiv\partial_a\Pi^a_r=0
\end{equation}
\begin{equation}
\phi\equiv\epsilon^{ab}\partial_b\left[\frac{\partial_aX^MP_M+\Pi^c_rF^r_{ac}}{\sqrt{W}}\right]=0
\label{intconstrabc}
\end{equation}
$\phi$ being the generator of the area preserving diffeomorphism.
There is also a global constraint arising from the elimination of
$X^{-}$, this is
\begin{equation}
\oint_{c}\left(\frac{\partial_aX^MP_M+\Pi^b_{r}F^r_{ab}}{\sqrt{W}}\right)d\xi^a=2
\pi n_c\label{intconstr2}
\end{equation}
where $c$ is a basis of homology of dimension one. $n_c$ are
integers associated to $c$.

$\beta_{ab}$ is the auxiliarly metric satisfying
\begin{equation}
\beta _{ab}=\left( 1-\beta^{-1}\Pi^a _{r}\Pi^b
_{r}\beta_{ab}\right) ^{-1}\left(
\partial _{a}X^{M}\partial _{b}X_{M}-\beta ^{-1}\Pi^c _{r}\Pi^d _{r}\beta_{ca}\beta_{db}\right)
\label{ecmovbeta}
\end{equation}
where $\beta$ is the determinant of the matrix $\beta_{ab}$.

$P_M$ are the conjugate momenta associated to $X^M$. The index $M$
refer to the transverse coordinates in the light-cone
decomposition of the target space. (\ref{intconstrabc}) and
(\ref{intconstr2}) arise from the integrability condition on the
resolution for $X^{-}$ and the further assumption that $X^-$,
winds up over $S^1$ with winding numbers ${n_c}$.

It is interesting to notice that the hamiltonian density
(\ref{hamil}) depends on the auxiliarly metric only through its
determinant $\beta$. In fact
\be
\b F^r_{ab}F^r_{cd}\b^{ac}\b^{bd}=\frac{1}{2}W(^*F^r)^2
\ee
where
\be
^*F^r\equiv\frac{\e^{ab}}{\sqrt{W}}F^r_{ab} \ee is the Hodge dual to the curvature
2-form $F^r$.

The determinant $\b$ may be obtained from (\ref{ecmovbeta}) after some calculations,
it has the following expressions
\be
\b=det(\p_aX^M\p_bX_M);\hspace{1.5cm}M=1,...,9 \ee for the $q=0$
case \cite{dwln},
\be \b=det(\p_aX^M\p_bX_M)+(\Pi^a\p_aX^M)^2;\hspace{1.5cm}M=1,...,8 \label{deb1} \ee for
the $q=1$ case and
\be
\b=det(\p_aX^M\p_bX_M)+(\Pi^a_r\p_aX^M)^2+
\frac{1}{4}(\Pi^a_r\Pi^b_s\e_{ab}\e^{rs})^2;\hspace{0.2cm}r=1,2;\hspace{0.2cm}M=1,...,7
\label{deb2} \ee for the $q=2$ case.

The hamiltonian densities obtained after replacing (\ref{deb1})
and (\ref{deb2}) into (\ref{hamil}) may also be constructed in a
more direct way from the hamiltonian density of the supermembrane
in the LCG by using duality in the canonical approach directly
without starting from the covariant formulation. Let us analyze
briefly this point. We consider the canonical action of the
supermembrane in the LCG \cite{dwln} with target space
$M_9$x$S^1$x$S^1$, its bosonic part is \be
{\cal H}_{SM}=\frac{1}{2}\frac{1}{\sqrt{W}}\left(
P^{M}P_{M}+det(\p_aX^M\p_bX_M)\right)+
\Lambda\e^{ab}\p_b\left(\frac{\p_aX^MP_M}{\sqrt{W}}\right);M=1,..,9
\label{Hamil2} \ee Consider, first, one of the compactified
coordinates taking values over $S^1$. It must satisfy
\begin{equation}
\oint_{c}dX=2\pi n_{c} \label{local}
\end{equation}
The terms involving that map in the canonical action are
\be
\langle P\dot{X}-\frac{1}{2}\frac{1}{\sqrt{W}}\left(P^2+
\e^{ac}\e^{bd}\p_aX\p_bX\p_cX^N\p_dX^N\right)+
\p_b\Lambda\e^{ab}\p_aX.\frac{P}{\sqrt{W}}\rangle\label{ac} \ee
where $X^N$ is different from $X$. We may then construct an
equivalent constrained term
\be
\langle PL_0-\frac{1}{2}\frac{1}{\sqrt{W}}\left(P^2+
\e^{ac}\e^{bd}L_aL_b\p_cX^N\p_dX^N\right)+
\p_b\Lambda\e^{ab}L_a\frac{P}{\sqrt{W}}\rangle \ee subject to
\be
\e^{ca}\p_cL_a=0;\hspace{1.5cm}\p_aL_0-\p_0L_a=0\label{const} \ee
We may introduce them into the action (\ref{ac}) through the use
of Lagrange multipliers, which we will denote $A_0$ and
$\e^{ab}A_b$ respectively. We then recognize that the conjugate
momenta to $A_b$ is
\be
\Pi^b\equiv\e^{ab}L_a \ee After the elimination of $L_0$ we get
\be
p=\e^{ab}\p_aA_b \ee (\ref{ac}) subject to (\ref{const}) reduces
then to
\be
\langle
\Pi^b\dot{A}_b-\frac{1}{2}\frac{1}{\sqrt{W}}\left[(\e^{ab}\p_aA_b)^2+
\Pi^c\Pi^d\p_cX^N\p_dX_N\right]-A_0\p_c\Pi^c-
\Lambda\Pi^b\p_b\left(\frac{\e^{cd}\p_cA_d}{\sqrt{W}}\right)\rangle\label{acci}
\ee

The terms (\ref{acci}) contribute together with the terms
independent of $X$ and $P$ in (\ref{Hamil2}) to give exactly the
same expression of the hamiltonian density (\ref{hamil}) and
(\ref{deb1}): \bea {\cal
H}_D=&&\frac{1}{2}\frac{1}{\sqrt{W}}\left(
P^{M}P_{M}+det(\p_aX^M\p_bX_M)+(\Pi^a\p_aX^M)^2+\frac{1}{4}W(^*F)^2\right)\nonumber\\&&
-A_0\p_c\Pi^c+\Lambda\e^{ab}\p_b\left(\frac{\p_aX^MP_M+\Pi^cF_{ac}}{\sqrt{W}}\right);\hspace{1
cm}M=1,...,8 \label{Hamil3} \eea (\ref{Hamil3}) is also the
hamiltonian density arising from the canonical formulation of the
Born-Infeld action \cite{arme}, it describes then the $D2-$brane
in 10 dimensions in the case of open supermembranes. If we now
repeat the above procedure for the second compactified coordinate
we obtain the following hamiltonian density \bea {\cal
H}=&&\frac{1}{2}\frac{1}{\sqrt{W}}\left(
P^{M}P_{M}+det(\p_aX^M\p_bX_M)+(\Pi^a_r\p_aX^M)^2+\frac{1}{4}(\Pi^a_r\Pi^b_s\e_{ab}\e^{rs})^2\right.\nonumber\\&&
\left.+\frac{1}{4}W(^*F^r)^2\right)
-A^r_0\p_c\Pi^c_r+\Lambda\e^{ab}\p_b\left(\frac{\p_aX^MP_M+\Pi^c_rF^r_{ac}}{\sqrt{W}}\right)
\label{Hamil4} \eea in complete agreement with (\ref{hamil}) and
(\ref{deb2}) which were obtained from the canonical analysis of
the covariant formulation of the theory. The equivalence between
${\cal H}_{SM}$, ${\cal H}_D$ and ${\cal H}$ may be then
established from the duality equivalence between the covariant
formulations of the theories, or more directly from the duality
equivalence of the gauge fixed canonical formulations in the LCG.
The relation becomes non-trivial because the procedure of going
from the covariant formulation to the LCG one involves the
elimination of the auxiliarly metric which is an on-shell step
while the duality equivalence are off-shell ones, they can be
formally performed on the functional integral.

\section{ The Minimal Configurations of the Hamiltonian}
We will now analyze more in detail (\ref{Hamil4}). Its supersymmetric extension may be
obtained in an straightforward way from the supermembrane
hamiltonian in the LCG by the procedure described above, we will
write the resulting expression at the end of the analysis. We may
solve explicitly the constraints on $\Pi^c_r$ obtaining
\be
\Pi^c_r=\e^{cb}\p_b\Pi_r;\hspace{1.2cm}r=1,2
 \ee
Defining the 2-form $\omega$ in terms of $\Pi_r$ as 
 
 \be
\omega  = \p_a \Pi_r \p_b \Pi_s\e^{rs} d\xi \wedge  d\xi', \ee
the condition of non trivial membrane winding imposes a restriction on 
it,  namely
\begin{equation}
\oint_{\Sigma } \omega =2 \pi n .
\end{equation}

With this condition on $\omega $, Weil's theorem ensures that there always exist an associated $U(1)$ principal bundle over 
$\Sigma $ and a connection on it such that $\omega $ is its curvature. 
The minimal configurations for the hamiltonian (\ref{Hamil4}) may be 
expressed in terms of such connections. 

In \cite{mor} the minimal configurations of the hamiltonian of the
double compactified supermembrane were obtained. In spite of the
fact that the explicit expression (\ref{Hamil4}) was not then obtained,
all the minimal configurations were found. They correspond to 
$\Pi_{r}$ = $\hat{\Pi}_{r}$ satisfying
\be
^* \hat{\omega} = \e^{ab}\p_a\hat{\Pi}_r\p_b\hat{\Pi}_s\e^{rs}=n\sqrt{W}\hspace{1cm}n\neq0 \ee 

The explicit expressions for $\hat{\Pi}_r$ were obtained in that paper
\cite{mor}. As mentioned before, they correspond to $U(1)$ connections on non
trivial principle bundles over $\Sigma$. The principle bundle is
characterized by the integer $n$ corresponding to an irreducible
winding of the supermembrane \cite{mrt}. Moreover the
semiclassical approximation of the hamiltonian density around the
minimal configuration, was shown to agree with the hamiltonian
density of super Maxwell theory on the world sheet, minimally
coupled to the seven scalar fields representing the coordinates
transverse to the world volume of the super-brane.

\section{The Symplectic Non-Commutative Formulation}

Let us now analyze the geometrical structure of the constructed hamiltonian.
We notice that the minimal configurations of the hamiltonian introduce 
a natural symplectic structure in the theory through the non 
degenerate 2-form $ \hat{\omega} $,
\be
\hat{\omega} = \p_a\hat{\Pi}_r\p_b\hat{\Pi}_s\e^{rs} d\xi \wedge  d\xi' \ee

Also, that $\hat{\Pi}^a_r$ is an invertible matrix. It allows
to define the metric $W_{ab}$ on the world volume.
\be
W_{ab}=2\p_a\hat{\Pi}_r\p_b\hat{\Pi}_r \ee Its determinant takes the value
\be
detW_{ab}=n^2W,\ee and its inverse is given by
\be
n^2W^{ab}=\frac{\e^{ac}}{\sqrt{W}}\frac{\e^{bd}}{\sqrt{W}}W_{cd}=
\frac{2}{\sqrt{W}}\hat{\Pi}^a_r\hat{\Pi}^b_r \ee

Furthermore, we introduce the covariant derivative $D_a$ with
respect to this metric $W_{ab}$, it then follows that
 \be
D_aW=0;\hspace{1cm}D_a\hat{\Pi}^b_r=0
 \ee
We now define the rotated covariant derivatives in terms of tangent 
space coordinates in the compactified directions:
\be
D_r\equiv\frac{\hat{\Pi}^a_r}{\sqrt{W}}D_a. \ee 
We may now perform a canonical transformation in order to introduce a 
symplectic connection ${\cal A}_r$in our formalism. The kinetic term
\be
\langle \Pi^a_r\dot{A}^r_a\rangle
\ee
may then be rewritten as
\be
\langle \Pi^{a}_r\dot{A}^{r}_a\rangle= \langle\e^{ab}\p_bA^{r}_a\dot{\Pi}_r\rangle=
\langle {\it \Pi^r}\dot{{\cal A}}_r\rangle \ee where we have introduced
\be
{\it \Pi^r}\equiv\e^{ab}\p_bA^r_a \ee
\be
{\cal A}_r\equiv\Pi_r-{\cal C}_r \label{newa}\ee where ${\cal C}_r$ is a time
independent geometrical object, which will be defined shortly. They satisfy the
following Poisson bracket relation
\be
\{{\cal A}_r(\xi),{\it \Pi}^{r}(\xi')\}_P=\d(\x,\x'). \ee 
The symplectic non commutative derivative ${\cal D}_r$ may be defined 
now as
\be
{\cal D}_r\equiv D_r+\{{\cal A}_r,\;\;\} \ee where the bracket
$\{\bullet ,\bullet \}$ is defined as follows
\be
\{\bullet ,\bullet \} \equiv\frac{2\e^{sr}}{n}D_r \bullet D_s \bullet 
=\frac{\e^{ba}}{\sqrt{W}}D_a \bullet D_b \bullet ;\hspace{1.4cm}n\neq0
\ee

We remark that these  symplectic non commutative derivatives  
behave as symplectic connections on a symplectic fibration over 
$\Sigma$ with the phase space $(X^{M}, P^{M})(\xi)$ being the fibre.
The gauge transformations generated by the first class constraint 
(area preserving diffeomorphisms in the base manifold $\Sigma $) 
preserve the Poisson bracket in the fibre . The symplectic non 
commutative derivatives preserve, in turn, the same structure, i.e the 
symplectic non commutative derivatives of the fields transform under 
gauge transformations in the same way as the fields and the 
holonomies generated by the symplectic connections preserve the 
Poisson bracket in the fibre. These properties may be checked out by straightforward 
calculations. In particular, ${\delta {{\cal A}_r}} $ = ${{{\cal D}_r} \xi}$ 
under infinitesimal gauge transformations with parameter $\xi$.    

Without loss of generality we rewrite (\ref{newa}) as
\be
\Pi^r={\cal A}_r+\hat{\Pi}_r \label{fa}\ee We then have for the
terms in (\ref{Hamil4}), \bea
\frac{1}{\sqrt{W}}\Pi^a_r\p_aX^M=&&\frac{1}{\sqrt{W}}\hat{\Pi}^a_r\p_aX^M
+\frac{\e^{ab}}{\sqrt{W}}\p_b{\cal A}_r\p_aX^M \nonumber\\ =&&
D_rX^M+\{{\cal A}_r,X^M\}={\cal D}_rX^M,\eea
\be
det\p_aX^M\p_bX_M=\frac{1}{2}\p_aX^M\p_cX^N\p_bX_M\p_dX_N\e^{ac}\e^{bd}
=\frac{1}{2}W\{X^M,X^N\}^2, \ee \bea
\Pi^a_r\Pi^b_s\e_{ab}\e^{rs}=&&n\sqrt{W}-2\sqrt{W}D_r{\cal
A}_s\e^{rs}+\e^{bc}\p_b{\cal A}_r\p_c{\cal A}_s\e^{rs}\nonumber\\
=&&n\sqrt{W}-\e^{rs}\sqrt{W}\left(D_r{\cal A}_s-D_s{\cal
A}_r+\{{\cal A}_r,{\cal A }_s\}\right)\nonumber\\=&&(n-{\cal
^*F})\sqrt{W},\eea where \bea {\cal ^*F}\equiv &&\e^{rs}{\cal
F}_{rs}\nonumber\\{\cal F}_{rs}\equiv&&D_r{\cal A}_s-D_s{\cal
A}_r+\{{\cal A}_r,{\cal A}_s\} \eea 
Finally, the generator of area preserving
diffeomorphisms
\be
\ph\equiv\e^{ab}\p_b\left(\frac{\p_aX^MP_M+\Pi^c_rF^r_{ac}}{\sqrt{W}}\right) \ee may
be expressed as
\be
-\ph={\cal D}_r{\it \Pi}^r+\{X^M,P_M\} \ee The hamiltonian density
(\ref{Hamil4}) may then be rewritten \bea H=\int_\Sigma {\cal
H}=\int_\Sigma\frac{1}{2\sqrt{W}}\left[(P^M)^2+({\it
\Pi^r})^2+\frac{1}{2}W\{X^M,X^N\}^2+W({\cal
D}_rX^M)^2\right.\nonumber
\\
\left.+\frac{1}{2}W({\cal F}_{rs})^2 \right]+
\int_\Sigma\left[\frac{1}{8}\sqrt{W}n^2-\Lambda\left({\cal
D}_r{\it \Pi}^r+\{X^M,P_M\}\right)\right]\hspace{1cm}\label{kht}
\eea
where the following global condition has been imposed
\be
\int_\Sigma {\cal ^*F}\sqrt{W}d^2\x=0\ee

The hamiltonian (\ref{kht}) may be extended to include the
fermionic terms of the supersymmetric theory. They may be obtained
from the hamiltonian of the supermembrane in \cite{dwln} by the
dual approach discussed previously. They are
\be
\int_\Sigma\sqrt{W}\left(\Lambda\{\bar{\theta}\Gamma_-,\theta\}-\bar{\theta}\Gamma_-\Gamma_r{\cal
D}_r\theta+\bar{\theta}\Gamma_-\Gamma_M\{X^M,\theta\} \right) \ee
where $\theta$ is the Majorana spinor of the original formulation
of the supermembrane in the LCG in D=11 which may be decomposed in
terms of a complex 8-component spinor of $SO(7)$x$U(1)$.

The hamiltonian (\ref{kht}) corresponds then exactly to a symplectic non-commutative
super-Maxwell theory on the world volume minimally coupled to
seven scalar fields $X^M$, $M=1,...,7$. The generator of area
preserving diffeomorphisms becomes the generator of gauge
transformations. In distinction to the star product defined in
\cite{SW} which depends on a constant large background
antisymmetric field of the string which couples to the $U(1)$
gauge fields of the D-brane, the symplectic non commutative product here is
intrinsically constructed from the minimal configurations of the
hamiltonian density which are unique (up to closed 1-forms) for each
given $n$ and related to the natural symplectic structure of the
world volume Riemann surface. This theory may be interpreted 
geometrically as a symplectic fibration over a Riemann surface, with 
fibre given by the symplectic phase space manifold generated by the 
transverse coordinate to the brane in the target space. Its symplectic 
structure being preserved under symplectomorphism induced by the first 
class constraint of the theory. The connection ${\cal D}_r $ is a 
symplectic connection on this symplectic fibration, i.e. the 
associated holonomies preserve the symplectic structure in the fibres 
\cite{Guille}. Whether this symplectic fibration with a symplectic 
connection could be globally extended in a consistent manner to  a 
type of Moyal non commutative gauge theory is an open question. As commented before, one can always 
globally deform the Poisson bracket in the fibration base space to a 
Moyal bracket, but it is not necessarily true that the symplectic 
structure on the fibre could be extended in the same way and, moreover, be 
preserved under holonomies.       

\section{Conclusions}
We have formulated the double compactified $D=11$ supermembrane dual with
non trivial irreducible winding as a symplectic non commutative Super Maxwell 
theory, i.e as an exact symplectic fibration over a compact Riemann Surface 
with a symplectic connection. The connection dynamics being governed by a 
hamiltonian that resembles that of a Maxwell theory. We emphasize 
that our construction is globally defined. Also, we remark that  the symplectic non-commutative gauge
theory we have introduced relies on the nonsingular minimal
configuration of the hamiltonian (\ref{Hamil4}), where the
assumption $n\neq0$ is essential. The minimal configuration
obtained in \cite{mor} correspond to the monopole connection
1-forms over Riemann surfaces \cite{mr} which may also be obtained
from a suitable pullback to $\Sigma$ of the connection
1-forms on the Hopf fibring over $CP_n$ \cite{pr}. Its curvature
is a non degenerate closed 2-form defining a natural symplectic
structure over $\Sigma$. The equivalence between the hamiltonian
(\ref{Hamil4}) of the double compactified D=11 supermembrane dual
and the hamiltonian (\ref{kht}) of the symplectic non-commutative geometry is
exact.

\acknowledgements 
I. Martin and A. Restuccia thank the kind 
hospitality of the Imperial College's Theoretical Group and the King's 
College Mathematics Department, respectively, where part of this 
work was done.


\begin{thebibliography}{99}
\bibitem{con} A. Connes, M. R. Douglas, and A. Schwarz, JHEP {\bf 9802:003} (1998), hep-th/9711162. M. R.
\bibitem{dh}Douglas and C. Hull, JHEP {\bf 9802:008,1998}, hep-th/9711165.%
\bibitem{ck}{Y.-K. E. Cheung and M. Krogh,  Nucl. Phys. {\bf B528}
(1998) 185.}%
\bibitem{ch}{C.-S. Chu and P.-M. Ho, Nucl. Phys. {\bf B550} (1999) 151, hep/th9812219; ``Constrained quantization of open string in
background B field and noncommutative $D$-brane,'' hep-th/9906192.}
\bibitem{vs}{V. Schomerus, JHEP {\bf 9906:030} (1999), hep-th/9903205.}%
\bibitem{hof}{C. Hofman and E. Verlinde,  JHEP {\bf 9812:010} (1998), hep-th/9810116; Nucl. Phys. {\bf B547} (1999) 157, hep-th/9810219.}%
\bibitem{pio}{B. Pioline and A. Schwarz, ``Morita Equivalence and
$T$-duality (or $B$ versus $\Theta$,'' hep-th/9908019.}%
\bibitem{howu}{P.~Ho and Y.~Wu, Phys. Lett. {\bf B398} (1997) 52, hep-th/9611233; P.-M.~Ho, Y.-Y. Wu and Y.-S.~Wu,  Phys. Rev. {\bf D58}
(1998) 026006, hep-th/9712201; P.~Ho, Phys. Lett. {\bf B434} (1998) 41, hep-th/9803166; P.~Ho and Y.~Wu,
 Phys. Rev. {\bf D58} (1998) 066003, hep-th/9801147}%
\bibitem{mli}{M.~Li, ``Comments on Supersymmetric Yang-Mills Theory on a
Noncommutative Torus,'' hep-th/9802052.}%
\bibitem{ko}{T.~Kawano and K.~Okuyama, Phys. Lett. {\bf B433} (1998) 29,
hep-th/9803044.}%
\bibitem{lizzi}{F.~Lizzi and R.J.~Szabo, ``Noncommutative Geometry and Space-time Gauge
Symmetries of String Theory,'' hep-th/9712206; G.~Landi, F.~Lizzi and R.J.~Szabo,
``String Geometry and the Noncommutative Torus,'' hep-th/9806099; F.~Lizzi and
R.J.~Szabo,``Noncommutative Geometry and String Duality,'' hep-th/9904064.}%
\bibitem{bigatti}{D.~Bigatti, Phys. Lett. {\bf B451} (1999) 324, hep-th/9804120.}%
\bibitem{bigsuss}{D.~Bigatti and L.~Susskind, ``Magnetic Fields, Branes
and Noncommutative Geometry,'' hep-th/9908056.}%
\bibitem{hashitz}{A.~Hashimoto and N.~Itzhaki, ``Noncommutative Yang-Mills
and the AdS / CFT correspondence,'' hep-th/9907166.}%
\bibitem{schi} L. Cornalba and R. Schiappa ``Matrix Theory Star Products from the Born-Infeld Action", hep-th/9907211.
\bibitem{SW} N. Seiberg and E. Witten ``String Theory and Noncommutative Geome\-try'', hep-th/9908142
\bibitem{Jabb} M.M. Sheikh-Jabbari''Noncommutative open string theories 
and their dualities'',hep-th/0101045.
\bibitem{Matsuo} Y. Matsuo, Y. Shibusa ``Volume preserving diffeomorphism and 
noncommutative branes'', hep-th/0010040.
\bibitem{Rudy} I. Rudychev ``From noncommutative string / membrane to 
ordinary ones'',hep-th/0101039. 
\bibitem{Ohta} K. Ohta, hep-th/0101082.
\bibitem{Amor} R. Amorim, J. Barcelos-Neto, hep-th/0101196.
\bibitem{giro} H. Girotti,M.Gomes, V.Rivelles, A.J. da Silva,Nucl. Phys. {\bf B587}(2000) 299. 
\bibitem{BST} E. Bergshoeff, E. Sezgin and P.K. Townsend, Phys. Lett. {\bf 189B} (1987) 75.
\bibitem{PKT95} P.K. Townsend, Phys. Lett. {\bf B350} (1995) 184, hep-th/9501068, Phys. Lett. {\bf B373} (1996) 68, hep-th/9512062.
\bibitem{mrt} I. Mart\'{\i}n, A. Restuccia and R. Torrealba, Nucl. Phys. {\bf B521} (1998) 117.
\bibitem{mor} I. Mart\'{\i}n, J. Ovalle and A. Restuccia  Phys. Lett. {\bf B472} (2000) 77.
\bibitem{vey-kont} J. Vey , Comment. Math. Helv. {\bf 50} (1975), 421; 
M. Kontsevich, ``Deformation quantization of Poisson manifolds'' 
q-alg/9709040.
\bibitem{dwln} B. de Wit, M. L\"usher and H. Nicolai, Nucl. Phys. {\bf B320} (1989) 135.
\bibitem{dwpp} B. de Wit, K. Peeters and J.C. Plefka, Nucl. Phys. 
Proc.Suppl. {\bf 62} (1998) 405.
\bibitem{arme} R. Manvelyan, A. Melikyan and R. Mkrtchyan, Phys. Lett. {\bf B425} (1998) 277.
\bibitem{Guille} V. Guillemin, E. Lerman and S. Sternberg, 
\underline {Symplectic Fibrations and Multiplicity Diagrams}, 
Cambridge University Press (1997)
\bibitem{mr} I. Mart\'{\i}n and A. Restuccia, Lett. Math. Phys. {\bf 39} (4) (1997).
\bibitem{pr} H. Porta and L. Recht, J. of Math. Anal. and App., {\bf 118},(2) (1986)
547; Revista Matem\'atica Iber. {\bf 2} (1986) 397.

\end{thebibliography}
\end{document}